\documentclass[aps,pra,twocolumn,showpacs,floatfix,superscriptaddress]{revtex4}

\usepackage{dcolumn}                 
\usepackage{nicefrac}
\usepackage{amsmath}
\usepackage{amsfonts}
\usepackage{amssymb}
\usepackage{amsthm}
\usepackage{bm}

\usepackage{color}
\usepackage{graphics}
\usepackage{graphicx}
\usepackage{epsfig}

\usepackage{dcolumn}                 
\newcolumntype{w}[1]{D{.}{.}{#1}}

\newcommand{\dd}{\mathrm{d}}

\newcommand{\KK}{\mathrm{K}}

\newcommand{\addrMissouri}{Missouri University of Science and
Technology, Rolla, Missouri 65409-0640, USA}

\newcommand{\addrGaithersburg}{National Institute of Standards and Technology,
Gaithersburg, Maryland 20899-8420, USA}

\newcommand{\addrPoznan}{Faculty of Chemistry,
Adam Mickiewicz University, Grunwaldzka 6, 60-780 Pozna\'n, Poland}

\begin{document}

\title{Thermal Correction to the Molar Polarizability of a Boltzmann Gas}

\author{U. D. Jentschura}
\affiliation{\addrMissouri}

\author{M. Puchalski}
\affiliation{\addrMissouri}
\affiliation{\addrPoznan}

\author{P. J. Mohr}
\affiliation{\addrGaithersburg}

\date{\today}

\begin{abstract}
Metrology in atomic physics has been crucial for a number of advanced
determinations of fundamental constants.  In addition to very precise
frequency measurements, the molar polarizability of an atomic gas has
recently also been measured very accurately.  Part of the motivation for
the measurements is due to ongoing efforts to redefine the International
System of Units (SI) for which an accurate value of the Boltzmann
constant is needed. Here, we calculate the dominant shift of the molar
polarizability in an atomic gas due to thermal effects. It is given by
the relativistic correction to the dipole interaction, which emerges
when the probing electric field is Lorenz transformed into the rest
frame of the atoms that undergo thermal motion. While this effect is
small when compared to currently available experimental accuracy, the
relativistic correction to the dipole interaction is much larger than
the thermal shift of the polarizability induced by blackbody radiation.
\end{abstract}

\pacs{51.30.+i, 06.20.F-, 06.20.-f, 47.80.Fg, 12.20.Ds}
\maketitle

%
%
\section{Introduction}

Spectacular progress in frequency metrology of 
simple atoms such as hydrogen~\cite{NiEtAl2000,ArNeJuBi2010,PaEtAl2011}
and helium~\cite{vREtA2011} has led to advances in our understanding 
of fundamental constants~\cite{MoTaNe2008}, and of their conceivable 
variation with time~\cite{FiEtAl2004}.
However, transition frequencies are not the 
only quantities that can be measured accurately
using currently available experimental methods.
The (static) molar polarizability $A_\epsilon$ of the helium-4 atom
has been determined in Ref.~\cite{ScGaMaMo2007} as
\begin{equation}
\label{Aeps}
A_\epsilon = \frac{ \alpha_{\rm d} \, N_{\rm A} }{3 \, \epsilon_0}
           = 0.5172535(47) \, \frac{{\rm cm}^3}{{\rm mol}} \,,
\end{equation}
where $N_{\rm A}$ is the Avogadro constant, $\epsilon_0$ is the vacuum permittivity, and
$\alpha_{\rm d}$ is the static electric dipole polarizability of helium.

Recently, the topic of gas thermometry has received considerable
attention through efforts to accurately measure the Boltzmann constant
as the basis for a possible redefinition of the kelvin in the
International System of Units 
(SI)~\cite{CaEtAl2009,PiEtAl2009,GaEtAl2010,LeEtAl2010,SuEtAl2010,ZhEtAl2010,
FeEtAl2010,PiEtAl2011,SuEtAl2011}.
The kelvin can be defined by assigning an exact specified value to the
Boltzmann constant, and in order to move forward with the redefinition,
it is necessary to know the current measured value as accurately as
possible so the specified value is well chosen~\cite{BIPMkelvin}.

The molar polarizability of helium-4 is also known from theory, so an
experiment that measures polarizability can instead be interpreted as a
measurement of pressure or a determination of the Boltzmann constant
$k_{\rm B}$.  The principle of the measurement of $k_{\rm B}$ in~\cite{ScGaMaMo2007} is
as follows.  The refractive index $\epsilon_r$ of the 
helium gas is deduced by measuring
microwave resonance frequencies of a helium-filled quasispherical cavity
as a function of pressure and temperature.  The index of refraction is
related to the molar density $\rho$ of the helium and its molar
polarizability, by the Clausius--Mossotti equation
\begin{equation}
\frac{\epsilon_r - 1}{\epsilon_r + 2} \approx A_\epsilon \rho \,.
\end{equation}
In the evaluation of the measurement,
a theoretical correction is applied to this formula,
which is mainly due to the diamagnetic susceptibility of the helium 
[see Eq.~(1) of Ref.~\cite{ScGaMaMo2007}].
The refractive index $\epsilon_r$ thus determines the
product 
\begin{equation}
A_\epsilon \rho = \frac{\alpha_{\rm d}}{3\epsilon_0} \, N_{\rm A}\rho \,.
\end{equation}
Knowing $A_\epsilon$ from Eq.~\eqref{Aeps}, one can solve for $\rho$.
The Boltzmann constant $k_{\rm B}$ follows from the 
real gas equation [``virial equation of state of helium gas'',
see Eq.~(2) of Ref.~\cite{ScGaMaMo2007}].
This equation is approximated by the ideal gas 
equation $p \approx R \, T \, \rho$, where 
$R = k_{\rm B} N_{\rm A}$ is the molar gas constant.
A crucial point of the measurement~\cite{ScGaMaMo2007} is that 
the resonator is maintained within a few millikelvins of 
the triple point of water, which is defined to be $273.16 \, {\rm K}$ 
in the SI. Measuring the pressure, stabilizing $T$ 
and having determined $\rho$, one can finally solve for $R$
and $k_{\rm B}$, determining the Boltzmann constant.

As outlined, an accurate value of the atomic polarizability is a prerequisite for the
measurement of $k_{\rm B}$.  In a thermal bath, the atom is not only subjected
to the probing low-frequency microwave radiation, but also to thermal
blackbody radiation.  By definition, the atomic polarizability describes
a second-order process where one of the two probing photons is absorbed,
while the other photon is emitted by the atom.  Additional interactions
involve the absorption and emission of blackbody photons and require
fourth-order perturbation theory.  At room temperature ($T = 300\,\KK$),
the blackbody radiation correction amounts to a relative
shift~\cite{PuJeMo2011} of the molar polarizability of helium by $4.0
\times 10^{-18}$.  This relative shift is numerically small, and it
would be somewhat surprising if the dominant thermal shift of the molar
polarizability in an atomic gas at room temperature were as small as
this.

We thus analyze a further shift of the polarizability, here, which is due to
the relativistic correction to the dipole interaction due to the thermal motion
of the atoms.  In the current brief report, we use units with $\hbar = c =
\epsilon_0 = 1$.  Calculations are reported in Sec.~\ref{theory} and
conclusions are drawn in Sec.~\ref{conclu}.

%
%
\section{Calculation}
\label{theory}

It has been known for some time that the interaction of a
compound system with an external electromagnetic field 
receives a correction (``R\"{o}ntgen term'') when 
the atom moves with respect to the electromagnetic field.
The thermal motion of atoms in a typical atomic gas
at room temperature follows Boltzmann statistics because the 
scale of the interatomic interactions (van-der-Waals and 
Casimir-Polder) is long compared to the de Broglie wavelength 
of the moving atoms.
The well-known R\"{o}ntgen term follows from the relativistic 
analysis of the electromagnetic interaction of a compound 
system with an external electromagnetic field $\vec E$.

The interaction of a compound system with the field is described by the
interaction Hamiltonian
\begin{equation}
H_I = -\vec D \cdot \vec E \, ,  \qquad
D^i = \sum_{a=0}^N e_a \, x_a^i\,,
\end{equation}
where the summation index $a$ is over all constituent particles of the
system, with the value $a=0$ being reserved for the atomic nucleus. The
charge of the $a^{\rm th}$ particle is denoted as $e_a$. The total
number of particles in the compound system is $N$.  
The dipole polarizability of an
atom can be written as [Eq.~(4) of Ref.~\cite{LaDKJe2010pra}]
\begin{align}
& \alpha_{\rm d}(\omega) = \frac{e^2}{3} \sum_{i=1}^3 \sum_\pm 
\nonumber\\[0.7ex]
& \; \times \left<\Psi_0 \left| 
\left( \sum_{n=1}^N x^i_n \right)
\frac{1}{H - E_0 \pm \omega} 
\left( \sum_{n=1}^N x^i_n \right)
\right| \Psi_0 \right> \,,
\end{align}
where $| \Psi_0 \rangle$ is the atomic ground state.
Here, the sum over $n =1, \dots,N$ is over all the atomic 
electrons (the atomic nucleus is at the origin 
of the coordinate system).  Evidently, the dipole polarizability is
essentially the second-order dipole interaction.
For a spherically symmetric ground 
state, all Cartesian components $i=1,2,3$ 
contribute equally to the dynamic polarizability, and the factor of
$1/3$ results from integration over angles in each component of the 
the dipole matrix element.
For small frequencies $\omega \to 0$, the symmetric 
limit  $\pm \, \omega \to 0$ leads to the replacement
\begin{equation}
\sum_\pm \frac{1}{H - E_0 \pm \omega}
\to 2 \, \left( \frac{1}{H - E_0} \right)' 
\end{equation}
and
\begin{eqnarray}
\alpha_{\rm d}(\omega) &\to& \alpha_{\rm d}(0) \to ~\alpha_{\rm d} \,,
\end{eqnarray}
where we denote the reduced Green function that 
enters the static polarizability by a prime~\cite{SwDr1991b}.

For an atom in motion, as described in Ref.~\cite{Pa2007}, the Lorentz
boost modifies the dipole interaction to be
\begin{equation}
\label{HIprime}
H'_I = -\vec D \cdot 
\left[ \vec E + \frac{1}{M} \, \left( \vec \Pi \times \vec B \right)
- \frac{\vec \Pi}{2 M} \, 
\left( \frac{\vec \Pi}{M} \cdot \vec E \right) \right] \,.
\end{equation}
Here, $M = \sum_a m_a$ is the total mass of the compound 
system (atom), $\vec E$ and $\vec B$ are the external
electric and magnetic fields, respectively, 
and $\vec \Pi = \sum_a \vec p_a$ is the total momentum 
of the compound system. 
The term proportional to the magnetic 
field vanishes after angular averaging
over the directions of motion of the atoms.
For the term quadratic in $\vec \Pi$, 
the angular averaging leads to a factor $1/3$
in the effective dipole interaction, leading to the correction
\begin{align}
H'_I =& \; -\vec D \cdot \left[ \vec E - \frac{\vec \Pi}{2 M} \,
\left( \frac{\vec \Pi}{M} \cdot \vec E \right) \right]
\nonumber\\[2ex]
\to & \; 
-\vec D \cdot \vec E \, \left( 1 - \frac{\vec v^2}{6 c^2} \right) \,,
\label{eq:dipshift}
\end{align}
where the factor of $c$ is restored in the denominator

The magnitude of the dipole interaction correction has a simple physical
interpretation.  The dipole interaction of an atom is essentially the
energy shift due to an applied electric field.  In its rest frame, the
moving atom sees a boosted electric field, which after averaging over
directions of the velocity, yields a correction factor of
$\left(1+v^2/3c^2\right)$ to the dipole energy.  Transformation of the
dipole energy in the rest frame of the atom to the laboratory frame
yields an additional correction factor of $\left(1-v^2/2c^2\right)$ for
a net correction of $\left(1-v^2/6c^2\right)$ for the effective dipole
interaction as given in Eq.~(\ref{eq:dipshift}).  Evidently, the dipole
correction has not been examined in detail beyond the linear
interaction, so we apply the same argument to the effective dipole
polarizability.  In this case, the interaction is quadratic in the
electric field, so the correction factor is $\left(1+2v^2/3c^2\right)$
for the boosted field strength.  The transformation to the laboratory
frame is the same, so the net correction is $\left(1+v^2/6c^2\right)$.
With this factor, the effective dipole polarizability of the moving atom
is
\begin{equation} \alpha_{\rm d}^\prime = \alpha_{\rm d} \, \left( 1 +
\frac{v^2}{6c^2} \right) \, .  \end{equation}

We are now in a position to average over the thermal ensemble.
With $\beta = 1/(k_{\rm B} T)$, 
where $k_{\rm B}$ is the Boltzmann constant and $T$ the thermodynamic
temperature, the Boltzmann velocity distribution is 
\begin{equation} 
f(\vec v) = 
\left( \frac{\beta \, M}{2 \pi} \right)^{3/2} \, 
\exp\left( - \frac{\beta \, M \, \vec v^2}{2} \right) \,,
\end{equation}
so that 
\begin{equation} 
\int \dd^3 v \, f(\vec v) = 1 \,.
\end{equation} 
In accordance with the equipartition theorem,
we find
\begin{equation} 
\left< \vec v^2 \right> = \frac{3}{\beta \, M}
= \frac{3 \, k_{\rm B} \, T}{M} \,,
\end{equation} 
so that the correction to the polarizability is 
\begin{equation}
\alpha_{\rm d}^\prime \to 
\alpha_{\rm d} \left( 1 + \frac{ k_{\rm B} \, T}{2 M c^2} \right) \equiv
\alpha_{\rm d} \left( 1 + \delta \right)  \,,
\end{equation}
where the last expression serves as a definition of $\delta$.
For room temperature $T = 300 \, {\rm K}$ 
and helium atoms, using physical constants 
from Ref.~\cite{MoTaNe2008}, 
we have a relative shift of 
\begin{equation} 
\label{delta}
\delta = \frac{ k_{\rm B} \, T}{2 M c^2} 
= 3.47 \times 10^{-12} \,.
\end{equation}
This effect is still small when compared to the experimental accuracy 
reported in Ref.~\cite{ScGaMaMo2007}.
However, it turns out to be much larger than the 
shift of the polarizability due to blackbody radiation,
which was previously calculated in Ref.~\cite{PuJeMo2011}.

%
%
\section{Conclusions}
\label{conclu}

High-precision measurements of the molar polarizability of atoms in
gaseous environments have become important for the determination of
fundamental constants (e.g., the Boltzmann constant) and for pressure
and temperature metrology.  The thermal corrections to the molar
polarizability are of importance because they represent effects which
cannot be easily brought under experimental control and would require
difficult adjustments of the experiments unless they can be shown to be
negligible. 

For an atom {\em at rest}, immersed in a
thermal bath, the blackbody radiation correction to the
polarizability~\cite{PuJeMo2011} is due to a fourth-order interaction
with the electromagnetic field (two blackbody photon, two photons of the
probing field) and is numerically small.  However, the measurement of
the polarizability usually proceeds in a Boltzmann gas, where atoms
are in thermal motion.  In the current, short paper, we find that the
dominant thermal shift of the molar polarizability in the latter case is
due to the R\"{o}ntgen term, i.e., due to necessity of transforming the
probing electric field into the rest frame of the moving atom by a
Lorentz transformation and transforming the energy shift back into the
laboratory frame.  The corresponding shift is given in
Eq.~\eqref{delta} and amounts to $\delta = 3.47 \times 10^{-12}$ for
helium at room temperature.
%
%
\section*{Acknowledgments}

The authors acknowledge helpful conversations with M.~R.~Moldover and
B.~N.~Taylor. This work has been supported by the National Science
Foundation and by the National Institute of Standards and
Technology (precision measurement grant).

\end{document}